%% file: manuscript.tex
\documentclass[5p,11pt,a4paper]{elsarticle}

\usepackage{graphicx}% Include figure files
\usepackage{dcolumn}% Align table columns on decimal point
\usepackage{bm}% bold math
\usepackage{hyperref}% add hypertext capabilities
\usepackage[mathlines]{lineno}% Enable numbering of text and display math
\usepackage{xcolor}
\usepackage{enumitem}
\usepackage{amsmath}
\usepackage{microtype}
\journal{Computational Materials Science}
\begin{document}
%%%%%% Start of article front matter
%\begin{frontmatter}
%\preprint{APS/123-QED}
%\begin{fmbox}
%\dochead{Research}

\title{ Dislocation nucleation and dynamics models for plastic deformation of
single crystalline tungsten: The role of interatomic potentials}% Force line breaks with \\

\author[a]{F. J. Dom\'inguez-Guti\'errez\corref{author}}
%\email{Corresponding author: javier.dominguez@ncbj.gov.pl}
\cortext[author] {Corresponding author: javier.dominguez@ncbj.gov.pl}
%\affiliation{%
%NOMATEN Centre of Excellence, National Center for Nuclear Research, 
%05-400 Swierk/Otwock, Poland
% }%
 
\author[b]{P. Grigorev}
%\affiliation{Aix-Marseille Universite, CNRS, CINaM UMR 7325,
%Campus de Luminy, 13288 Marseille, France}
\author[a]{A. Naghdi}
\author[a]{Q. Q. Xu}
%\affiliation{%
%NOMATEN Centre of Excellence, National Center for Nuclear Research, 
%05-400 Swierk/Otwock, Poland
% }%
\author[c]{J. Byggm\"astar}
%\affiliation{Department of Physics, P.O. Box 43, FI-00014
%University of Helsinki, Finland}
\author[c,d]{G. Y. Wei}
%\affiliation{Department of Physics, P.O. Box 43, FI-00014
%University of Helsinki, Finland}
%\affiliation{Henan Academy of Big Data, Zhengzhou University, 
%Zhengzhou 450052, China}
\author[b]{T. D. Swinburne}
%\affiliation{Aix-Marseille Universite, CNRS, CINaM UMR 7325,
%Campus de Luminy, 13288 Marseille, France}
\author[a]{S. Papanikolaou}
%\affiliation{%
%NOMATEN Centre of Excellence, National Center for Nuclear Research, 
%05-400 Swierk/Otwock, Poland
% }%
\author[a]{M. J. Alava}
%\affiliation{%
%NOMATEN Centre of Excellence, National Center for Nuclear Research, 
%05-400 Swierk/Otwock, Poland
% }%
%\affiliation{Department of Applied Physics, Aalto University, P.O. Box 11000, 00076 Aalto, Espoo, Finland}
\address[a]{NOMATEN Centre of Excellence, National Centre for Nuclear Research, ul. A. Sołtana 7, 05-400 Otwock, Poland}
\address[b]{Aix-Marseille Universite, CNRS, CINaM UMR 7325,
Campus de Luminy, 13288 Marseille, France}
\address[c]{Department of Physics, P.O. Box 43, FI-00014
University of Helsinki, Finland}
\address[d]{Henan Academy of Big Data, Zhengzhou University, 
Zhengzhou 450052, China}
\address[e]{Department of Applied Physics, Aalto University, P.O. Box 11000, 00076 Aalto, Espoo, Finland}
%\end{fmbox}
\date{\today}% It is always \today, today,
%%%%%%%%%%% Abstract
%\begin{abstractbox}
\begin{abstract}

Computational modeling is usually applied to aid experimental 
exploration of advanced materials to better 
understand the fundamental plasticity mechanisms 
during mechanical testing.
In this work, we perform Molecular dynamics (MD) simulations
to emulate experimental room temperature spherical--nanoindentation 
of crystalline W matrices by different interatomic potentials: 
EAM, modified EAM, and a recently developed machine learned based
tabulated Gaussian approximation potential (tabGAP)
for describing the interaction of W-W. 
Results show similarities between load displacements 
and stress--strain curves, regardless of the numerical model.
However, a discrepancy is observed at early stages of the 
elastic to plastic deformation transition showing different 
mechanisms for dislocation nucleation and evolution, that is 
attributed to the difference of Burgers vector magnitudes, 
stacking fault and dislocation glide energies.
Besides, contact pressure is investigated by considering 
large indenters sizes that provides a detailed analysis of 
screw and edge dislocations during loading process.
Furthermore, the glide barrier of this kind of dislocations 
are reported for all the 
interatomic potentials showing that tabGAP model presents the 
most accurate 
results with respect to density functional theory calculations 
and a good qualitative 
agreement with reported experimental data.

\end{abstract}

%Use showkeys class option if keyword
%\end{abstractbox}
%\end{fmbox}
%\end{frontmatter}
\maketitle
%\linenumbers
%\tableofcontents
%%%%%%%%%%%%%%%%%%%%%%%%%%%%%%%%%%%%%%%%%%%%%%%%%%%%%%%%%%%
\input{sections/introduction}
\input{sections/methods}

\input{sections/results_nanoind}
\input{sections/dislocations}

\input{sections/sizeEffects}
\input{sections/concluding}

\section*{Acknowledgements}
%\ack
%We acknowledge 
We would like to thank M-C Marinica for inspiring
conversations.
We acknowledge support from the European Union Horizon 2020 research
 and innovation program under grant agreement no. 857470 and from the 
 European Regional Development Fund via the Foundation for Polish 
 Science International Research Agenda PLUS program grant 
 No. MAB PLUS/ 2018/8. We acknowledge the computational resources 
 provided by the High Performance Cluster at the National Centre 
 for Nuclear Research in Poland.
PG gratefully recognizes support from the Agence Nationale de Recherche, 
via the MeMoPAS project ANR-19-CE46-0006-1 as well as access to the HPC
resources of IDRIS under the allocation A0090910965 attributed by GENCI. 
 %and also the Seawulf institutional 
 %cluster at the Institute for Advanced Computational Science in 
 %Stony Brook University.
\section*{Data Availability Statement}
The raw data and MD simulations results and nanoindentation tests
of W sample visualization that support the findings of this study
are available within the article and its supplementary material.
\nocite{*}

%\appendix
%\input{sections/uniaxial}

\bibliography{bibliography}
\bibliographystyle{elsarticle-num}% Produces the bibliography via BibTeX.
%\bibliographystyle{vancouver}

%\begin{figure*}[b!]
%    \centering
%    \includegraphics[width=\textwidth]{supplementary/Supplemental_material_1.pdf}
%\end{figure*}
%\begin{figure*}[b!]
%    \centering
%    \includegraphics[width=\textwidth]{supplementary/Supplemental_material_2.pdf}
%\end{figure*}

\end{document}

%% file: sections/introduction.tex
\section{\label{sec:intro}Introduction}

The development of novel technology for current
energy industries requires the use of refractory 
BCC materials that can mechanically sustain extreme 
operating conditions 
that may include, among others, high temperature and
irradiation. 
Tungsten can fulfill these requirements and is the main 
candidate material to design a Plasma 
Facing Component (PFM) for fusion reactors
\cite{PITTS2013S48,RIETH2011463,PAMELA2009194,HOLTKAMP200998}
due to its high melting point, low sputtering yield, and low
tritium inventory 
\cite{mayer,PIAGGI2015233,Lipschultz,KAUFMANN2007521,SCHWARZSELINGER2018228}. 
Like the International Thermonuclear Experimental Reactor (ITER), 
the main objective is to develop accessible applications for W and its alloys 
in extreme operating environment and to identify limitations
in future fusion reactors. 
This will certainly increase the already demanding mechanical properties 
of the materials facing higher neutron irradiation dose and nuclear 
transmutation rates \cite{PITTS2013S48,RIETH2011463}, where 
computational models can save financial and technological resources 
\cite{1998562,WEN20173626,Wangexp}.

In a fusion reactor, W experiences a harsh environment due to the 
alpha particles and hydrogen ions irradiation that causes
indentation size effects. Thus, the modeling of nanoindentation 
of W is essential and of importance to better understand its
behaviour at a similar scale (e.g. within ~100 nm of the surface) 
\cite{BEAKE201863}.
Atomistic simulations based on the molecular dynamics (MD) method 
have been used to provide an insight of plastic deformation of BCC 
metals by spherical nanoindentation 
\cite{JavVarilla,SCHUH200632,Javier2021,KURPASKA2022110639,VARILLAS2017431,BEAKE201863,1998562,PO2016123}, 
as well as the description of the 
the physical and chemical processes for defect production 
during external load where the interatomic potentials play an 
important role to model the W-W interaction 
\cite{doi:10.1063/1.5094852,doi:10.1063/1.3298466,PhysRevB.100.144105,NatureKai,Marinica_2013,Wang_2017,GOEL2018196}.
For example, large scale MD simulations are mainly based on the 
embedded atom method (EAM) expressing the energy of an atomic 
system by the sum of pair potential terms of the separation between an 
atom and the collective chemical bonds between the neighboring atoms \cite{Marinica_2013}. 
This kind of approach provides information about the atomistic 
description of dislocation nucleation during mechanical loading, 
with some limitations in several cases.
The nucleation of a dislocation with a Burgers
vector $1/2 \langle 111 \rangle$ commonly occurs during 
nanomechanical tests of BCC metals like W and hence it is important 
to accurately model the glide of 1/2$\langle 111 \rangle$\{110\}
screw dislocations 
for describing dislocation dynamics numerically 
\cite{PhysRevMaterials.4.023601}, being the goal of our work.

Our paper is organized as follows: In Section 2 we
describe the details of the numerical simulations by 
considering different interatomic potentials based on 
EAM \cite{Wang_2017,Marinica_2013}, 
modified EAM \cite{HIREMATH2022111283}
and tabulated Gaussian 
approximation potential framework \cite{byggmastar_simple_2022}. 
In Section 3, the atomistic insights of the 
mechanical response of crystalline W by 
nanoindentation tests 
are presented on the [001], [011], 
and [111] orientations. 
We analyze dislocation density as function indentation 
depth, in conjunction to dislocation loop formation, 
pile-ups formation showing nanoindentation mechanisms. 
Stress-strain curves and dislocation densities 
are reported.
Besides, an analysis of the screw and junction dislocations
by NEB is presented to elucidate the difference on 
dislocation dynamics in the MD simulation due 
to the modeling of the interatomic potential. 
%Agreement with experimental measurements is reported and discussed, in
%connection to dislocation junction formation. 
Finally, in
section 4, we provide concluding remarks.

%% file: sections/methods.tex
\section{\label{sec:methods}Computational methods}

%\subsection{Interatomic potentials}

In general, MD simulations utilize EAM approach as a first attempt
to computationally model the physical and mechanical processes of
the nanomechanical response of crystalline W under external load, 
which defines the energy of the $i$--atom as
\begin{equation}
    E_i = \frac{1}{2} \sum_{j \in \mathcal{N}_i} V \left(r_{ij}\right) +
    F[\rho_i],
\end{equation}
where $\mathcal{N}_i$ represents the atoms within the cutoff range 
$d= 0.44$ nm, $V(r)$ is a pairwise potential, repulsive at short
range, and $F [\cdot]$ the embedding function for an atom in a region
of electron density, given by $\rho_i = \sum_{j \neq i} \phi(r_{ij})$.
Therefore, we explore the advantages and limitations of the following 
interatomic potentials over a traditional EAM model: % based on embedded atom method (EAM) formalism:

\begin{table*}[t!]
\centering
\caption{
Elastic constants and surface energies in ev/$\AA{}^2$ of 
tungsten obtained by MD simulations and compared with 
experimental values \cite{hertzberg1977deformation}, in units of GPa. 
%The lattice constant, $a_0$, used in MD calculations
%is 3.1472 
%\AA{} for EAM1 \cite{Marinica_2013} and 3.16 \AA{} for 
%EAM2 \cite{Wang_2017}
; with Burgers vector $\vec b = a_0\sqrt{3}/2$ and lattice constant $a_0$.}
\begin{tabular}{l rrrrrr}
\hline
Parameter & Wang & Marinica & TabGAP & Mason & Hiremath & Exp. \\
\hline
$C_{11}$  &  523   &   523  & 540 & 511 & 527 & 501 \\
$C_{12}$  &  203   &   203  & 182 & 201 & 194 & 198 \\
$C_{44}$  &  160   &   160  & 133 & 161 & 177 & 151 \\
%\hline
Shear Mod.   & 160.0  & 160.0 & 135 & 161 & 177 & 151.47 \\
Poisson R.   & 0.28   & 0.28 & 0.28  & 0.27 & 0.27 & 0.283 \\
Elastic Mod. & 409.48 & 409.48 & 448.63 & 431.03 & 424.63 & 388.82 \\
%\hline
%\multicolumn{4}{c}{Defect properties} \\
%\hline
%Value & Wang & Marinica & TabGAP & Mason & Hiremath & DFT \\
\hline
$E_{\rm coh}$ BCC (eV/atom) & -8.899 & -8.899 & -8.9 & -8.65 & -8.90 & -8.9  \\
%$E_{\rm f}^{<111>}$ (eV) & 10.53 & 10.53 & & & & 10.53 \\
%$E_{\rm f}^{<110>}$ (eV) & 10.82 & 10.82 & & & & 10.82 \\
%$E_{\rm f}^{V}$ (eV) & 3.49 & 3.49 & & & & 3.49 \\
$|b| = a_0 \sqrt{3}/2$ & 2.72 & 2.72 & 2.75 & 2.74 & 2.74 & 2.74 \\
\hline
\end{tabular}
\label{tab:EC_Mo} 
\end{table*}

\begin{description}[leftmargin=0cm]
%%%%%%%%%%%%%%%%%%%%%%%%%%%%%%%%%%%%%%%%%%%%%%%%%%%%%%%%%%%%%%%%%
\item[EAM1] 
This interatomic potential is the main reference of our work and 
was developed by Marinica et al. 
\cite{Marinica_2013} (second class of potentials defined as
EAM2 by the authors) for an investigation of radiation 
defects as well as dislocation nucleation in tungsten. 
This potential was fitted to:
1) experimental values such as lattice 
constant measurements, cohesive energies of tungsten 
in BCC phase and elastic constants; 
2) basic point defects formation energies calculated by 
\textit{ab--initio} computations for
the single vacancy and self-interstitials with 
different crystal orientations, namely 
$\langle 100 \rangle$, 
$\langle 110 \rangle$, and $\langle 111 \rangle$, 
interatomic forces for liquid state
configurations of tungsten. 
The resulting EAM potential has been evaluated by comparing
with DFT results on point
defects (I2 and I4 interstitial clusters as well as two-, 
three-- and four--vacancy clusters formation energy) as well
as extended defects (surface energy, Bain 
deformation energy, dislocation core energy and Peierls
energy barrier calculations). 
The results show that, in principle, this potential is a good 
choice for dislocation involved simulations of
tungsten, such as nanoindentation measurements.
%%%%%%%%%%%%%%%%%%%%%%%%%%%%%%%%%%%%%%%%%%%%%%%%%%%%%%%%%%%%%%%%%%%%%%
\item[EAM2] This EAM potential is reported by 
Wang et al. \cite{Wang_2017} where the W--W interaction is in 
principle based on Marinica et al potential 
\cite{Marinica_2013}, and 
developed by computing DFT calculation to fit
the W--H interactions taken from reported and generated 
W--W/W--H DFT data by the authors. 
Here, defect properties like hydrogen with SIA, 
hydrogen diffusion in strained 
W are reported. 
Thus, the W--W interaction description was optimized 
to include the W--H parametrization giving the opportunity to 
model W--H interaction for further 
investigation of the effect of deuterium on the dislocation 
nucleation and dynamics during mechanical testing.

%%%%%%%%%%%%%%%%%%%%%%%%%%%%%%%%%%%%%%%%%
\item[EAM+ZBL] Mason et al. \cite{Mason_2017} introduced
smoothly--varying, physically-- motivated modifications 
to the Ackland--Thetford potential~\cite{ackland_improved_1987} 
and adding  $V_{\rm ZBL}(r)$ the ZBL universal screening 
potential contribution to improve
vacancy- and surface-related properties. 
This potential was parametrized to perform 
simulations of vacancy--type defects for collision cascades at 
elevated temperatures with improved surface properties for pure W.

%%%%%%%%%%%%%%%%%%%%%%%%%%%%%%%%%%%%%%%%%%%%%%%%%%%%%%%%%%%%%%%%%

\item[TabGAP] The tabulated Gaussian approximation potential 
(GAP) was developed by Byggmästar et al. \cite{byggmastar_simple_2022}. 
It is a GAP machine--learning potential 
that has been trained with only simple low--dimensional 
descriptors (two--body, three--body,
and an EAM-like density). The low dimensionality of the 
descriptors allows for creating
faster tabulated versions, where the machine-learning energy 
contributions are mapped 
onto grids \cite{glielmo_efficient_2018,Vandermause,JesperTabGAP}.
The total energy is then evaluated efficiently using cubic splines as
\begin{equation}
\begin{aligned}
    & E_\mathrm{tot.} =  \sum_{i < j}^N S_{ij}^\mathrm{1D} (r_{ij}) + \sum_{i, j<k}^N S_{ijk}^\mathrm{3D} (r_{ij}, r_{ik}, \cos \theta_{ijk}) \\
    & + \sum_{i}^N S_\mathrm{emb.}^\mathrm{1D} \left(\sum_j^N S_\varphi^\mathrm{1D} (r_{ij}) \right).
\end{aligned}
\end{equation}
%\begin{equation}
%    E_\mathrm{tot.} =  \sum_{i < j}^N S_{\mathrm{rep. + 2b}}^\mathrm{1D} (r_{ij}) + %\sum_{i, j<k}^N S_{ijk}^\mathrm{3D} (r_{ij}, r_{ik}, \cos \theta_{ijk})
%     + \sum_{i}^N S_\mathrm{emb.}^\mathrm{1D} \left(\sum_j^N S_\varphi^\mathrm{1D} %(r_{ij}) \right),
%\end{equation}
Here, $S_{ij}^\mathrm{1D} (r_{ij})$ represents a one--dimensional cubic 
spline for the two-body contribution, $S_{ijk}^\mathrm{3D}$ is the 
three-- dimensional spline for the three--body contribution, and the final
term is the embedding energy contribution similar to the EAM potentials. 
Despite the simplicity compared to other machine--learning potentials, 
the tabGAP achieves meV/atom accuracy for tungsten--based high-entropy
alloys and compares well with DFT for various elastic, defect, 
and melting properties \cite{byggmastar_simple_2022,JesperTabGAP} 
that can be applied to model defect production at high temperatures 
\cite{DOMINGUEZGUTIERREZ202238}.

%%%%%%%%%%%%%%%%%%%%%%%%%%%%%%%%%%%%%%%%%%%%%%%%
\item[MEAM] Hiremath et al. \cite{HIREMATH2022111283} 
developed a second nearest-neighbor modified EAM potential 
to investigate mechanisms of fracture in W samples providing an atomic
insight; which also yields surface and unstable twinning energies
that are in slightly better agreement with DFT.
Here, the total potential energy of the system is given as
\begin{equation}
E = \sum_i F_i(\overline{\rho}_i) + \sum_i \sum_{j \neq i}
S_{ij}\phi_{ij}(r_{ij}),
\end{equation}
where $F_i(\overline{\rho}_i)$ represents the embedding energy 
associated with placing the $i$-atom into the background electron density 
$\overline{\rho}_i$. 
The function $\phi_{ij}(r_{ij})$ is defined as the pair 
interaction contribution between $i$ and $j$ atoms, separated
by the distance $r_{ij}$, while $S_{ij}$ is a screening function.
The fitting process was done by using the open-source M-EAM 
parameter calibration (MPC) tool \cite{MEAMCode} to reproduce 
DFT data that serves as input data.

\end{description}

\begin{table}[b!]
\centering
\caption{Surface energy in eV/$\AA{}^2$ as a function of miller 
index computed by all approaches. DFT values are taken from \cite{VITOS1998186}.}\label{tab:SEMD}
\begin{tabular}{lrrrr}
\hline
 Approach      & $\{001\}$ & $\{110\}$ & $\{111\}$ & $\{112\}$ \\
\hline
%DFT (Marinica et. al.)           & 0.32  & 0.25 & -- & 0.27 \\
DFT             & 0.29  & 0.25 & 0.28 & 0.26 \\
Wang et al.    & 0.17  & 0.14 & 0.18 & 0.17 \\
Marinica et al. & 0.17 & 0.14 & 0.18 & 0.17 \\
TabGAP         &  0.24 & 0.21 & 0.24 & 0.23  \\
Mason et al.   &  0.24 & 0.22 & 0.26 & 0.24  \\
Hiremath et al. & 0.24 & 0.21 & 0.25 & 0.24    \\
\hline
\end{tabular}
\end{table}

%%%%%%%%%%%%%%%%%%%%%%%%%%
%%%%%%%% Nanoindentation
%%%%%%%%%%%%%%%%%%%%%%%%%%%%%%%%%
\subsection{Nanoindentation test}
\label{subsec:nanoindentation}

\begin{table*}[t!]
\centering
\caption{Size of the numerical samples used to perform MD simulations.}\label{tab:MD_data}
\begin{tabular}{clllll}
\hline
Orientation   & Size(dx,dy,dz) [nm] & X-axis & Y-axis & Z-axis & Atoms  \\
\hline
$[001]$ & (37.92,41.08,31.60) & $(100)$  & $(010)$ & $(001)$ & 3 120 000 \\
$[011]$ & (34.76,37.99,36.65) & $(100)$  & $(01\overline{1})$ & (011) & 3 066 800 \\
$[111]$ & (33.52,34.83,46.52) & $(\overline{1}01)$ & $(1\overline{2}1)$& 
$(111)$ & 3 442 500 \\
\hline
\end{tabular}
\end{table*}

In order to perform the MD simulations, we utilize the Large-scale Atomic/ Molecular 
Massively Parallel Simulator (LAMMPS) software \cite{THOMPSON2022108171}.
To model 
correctly plastic deformation, we first compute the the elastic constants, 
C$_{ij}$, and other W properties, as well as the surface energies
by all the interatomic potentials considered in this work.
%, for a small BCC W sample of 3.375 nm$^3$;
%Interstitial energy formation on $\langle 111 \rangle$ and 
%$\langle 110 \rangle$, and 
%vacancy energy formation 
%that is obtained as:
%\begin{equation}
%    E^f_{i,v} = E_f - \left[ \frac{N_0 \pm 1}{N_0} \right]E_0,
%\end{equation}
%where $E^f$ is the energy formation of interstitial or vacancy; 
%$E_f$ is the final energy after energy optimization process by 
%considering the interstitial atom or vacancy in the computational 
%cell; $\pm$ is $+$ for interstitial energy and $-$ for 
%vacancy formation energy computation, respectively.
The obtained values are presented in Table~\ref{tab:EC_Mo} and 
\ref{tab:SEMD} noticing 
that the EAM-based potentials we utilize are similar in most respects.
%W is a metal that shears in $\langle 111 \rangle$ 
%directions on the $\{110\}, \{112\}$, and $\{123\}$ planes.

We apply MD simulations through an $NVE$ statistical 
thermodynamic ensemble and the velocity Verlet algorithm to emulate experimental
nanoindentation test. 
Periodic boundary conditions are set on the $x$ and $y$ axes to 
simulate an infinite surface, while the $z$ orientation contains 
a fixed bottom boundary and a free top boundary in all 
MD simulations \cite{DOMINGUEZGUTIERREZ2021141912}.
We first defined the initial W sample with respect to its crystal orientation as 
shown in Tab. \ref{tab:MD_data}
followed by a process of energy optimization and equilibration for
100 ps with a Langevin thermostat at 300 K and a time constant of
100 fs \cite{DOMINGUEZGUTIERREZ2021141912}. 
This is done until the system reaches a homogeneous sample 
temperature and pressure profile with a density of 19.35 g/cm$^3$, 
which is similar to the experimental value. 
At the first stage, the samples are defined into three sections 
in the $z$ direction for setting up boundary conditions along 
its depth, $dz$: 
1) frozen section with a width of $\sim$0.02$\times dz$ 
for stability of the numerical cell; 
2) a thermostatic section
at $\sim$0.08$\times dz$ above the frozen one to dissipate the 
generated heat during nanoindentation; and 3) the dynamical 
atoms section, where the interaction with the indenter tip modifies
the surface structure of the samples. 
In addition, a 5 nm vacuum section is included at the top 
of the sample \cite{KURPASKA2022110639}. 

The indenter tip is considered as a non-atomic
repulsive imaginary (RI) 
rigid sphere with a force potential defined as: 
$F(t) = K \left(\vec r(t) - R \right)^2$ where $K = 236$ 
eV/\AA$^3$ (37.8 GPa) is the force constant, and $\vec r(t)$ is 
the position of the center of the tip as a function of time,
with radius $R$. 
Here, $\vec r(t) = x_0 \hat x + y_0 \hat y + (z_0 \pm vt)\hat z$ 
with $x_0$ and $y_0$ as the center of the surface sample on the 
$xy$ plane, the $z_0 = 0.5$ nm is the initial gap between the surface 
and the intender tip moves with a speed $v$ = 20 m/s.
The loading and unloading processes are defined by considering 
the direction of the velocity as negative and positive, respectively.
Each process is performed for 125 ps with a time step of 
$\Delta t = 1$ fs. 
The maximum indentation depth is chosen to 3.0 nm 
to avoid the influence of boundary layers in the 
dynamical atoms region.

%% file: sections/results_nanoind.tex
\section{\label{sec:results}Results}

%\subsection{Nanoindentation}

\begin{figure}[!b]
   \centering
   \includegraphics[width=0.45\textwidth]{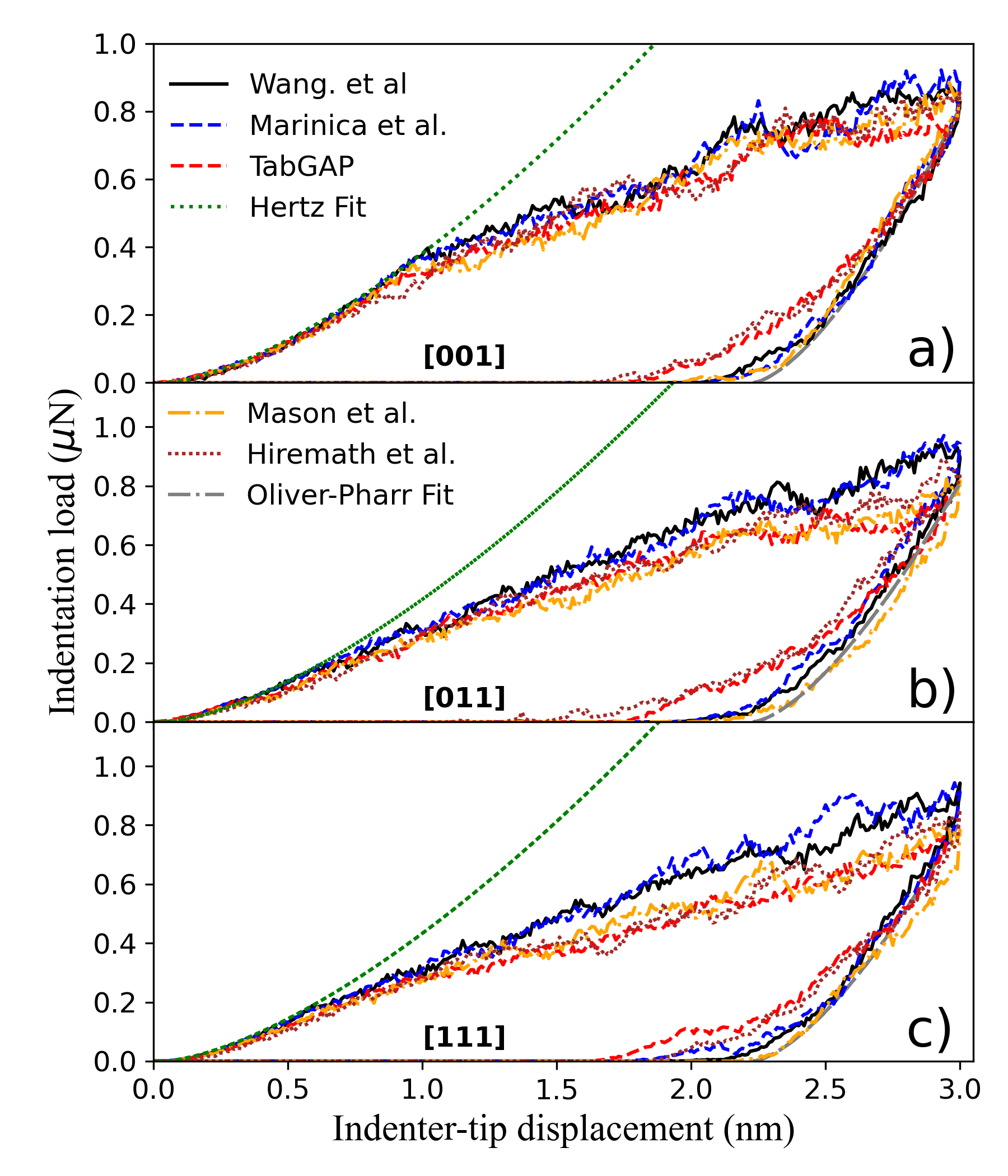}
   \caption{(Color on-line). Load displacement curves showing the 
   loading and unloading process during nanoindentation test in a). 
   Hertz and Oliver-Pharr fitting are added to show the 
   mechanical processes.
%   Indentation stress as a function of indentation strain obtained 
%   by MD simulation in b).
   }
   \label{fig:LDcurve}
\end{figure}

From our MD simulations, the loading and unloading processes of 
nanoindentation test of W sample are recorded and shown in Fig.
\ref{fig:LDcurve} by using different interatomic potentials and an indenter 
tip radius $R = 6$ nm. 
We include a Hertz fitting curve based on the sphere--flat 
surface contact and  expressed as 
\begin{equation}
P_{\rm H} = \frac{4}{3}
E_{\rm eff}R^{1/2}h^{3/2},
\end{equation}

\begin{figure*}[!t]
   \centering
   \includegraphics[width=0.3\textwidth]{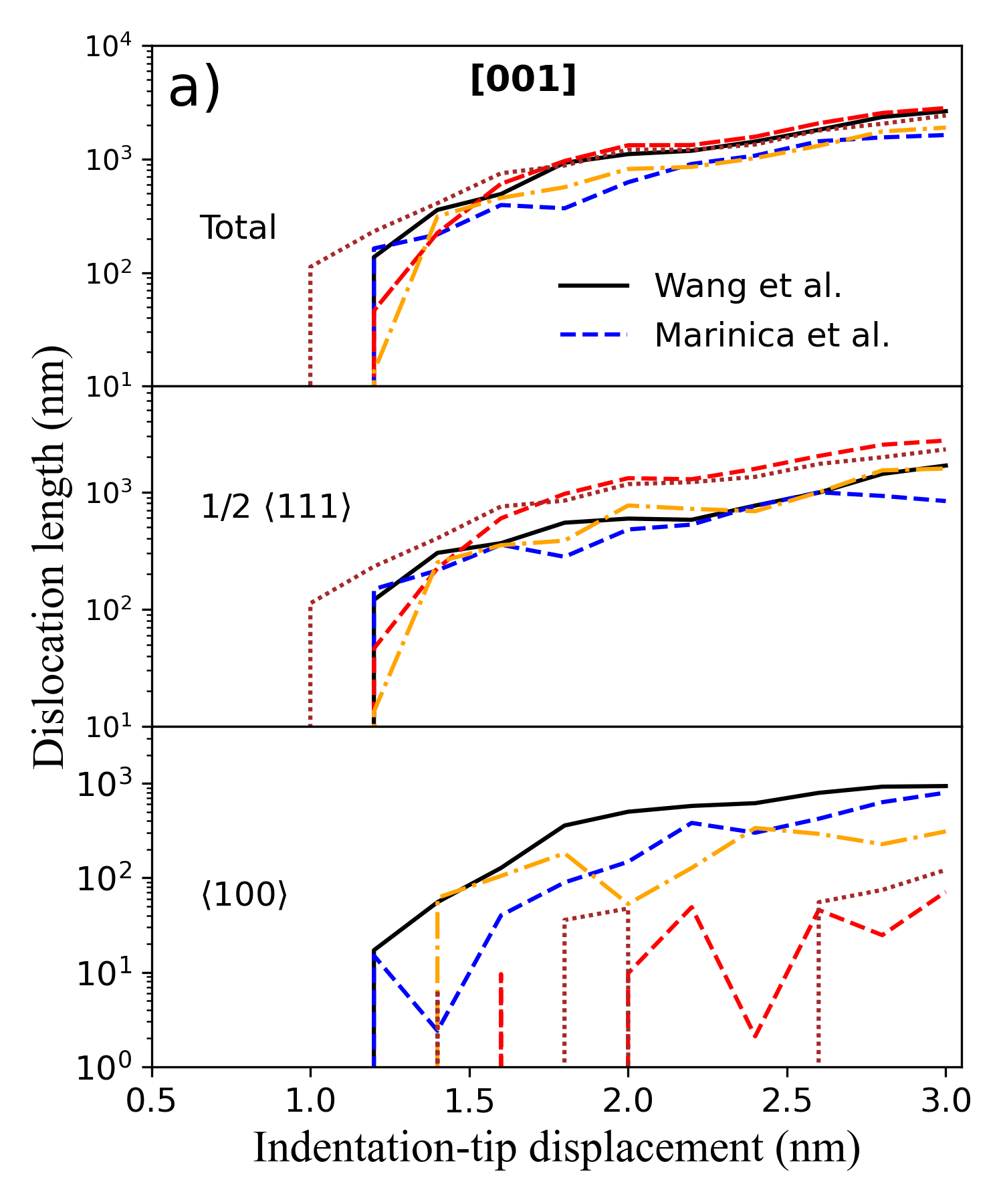}
   \includegraphics[width=0.3\textwidth]{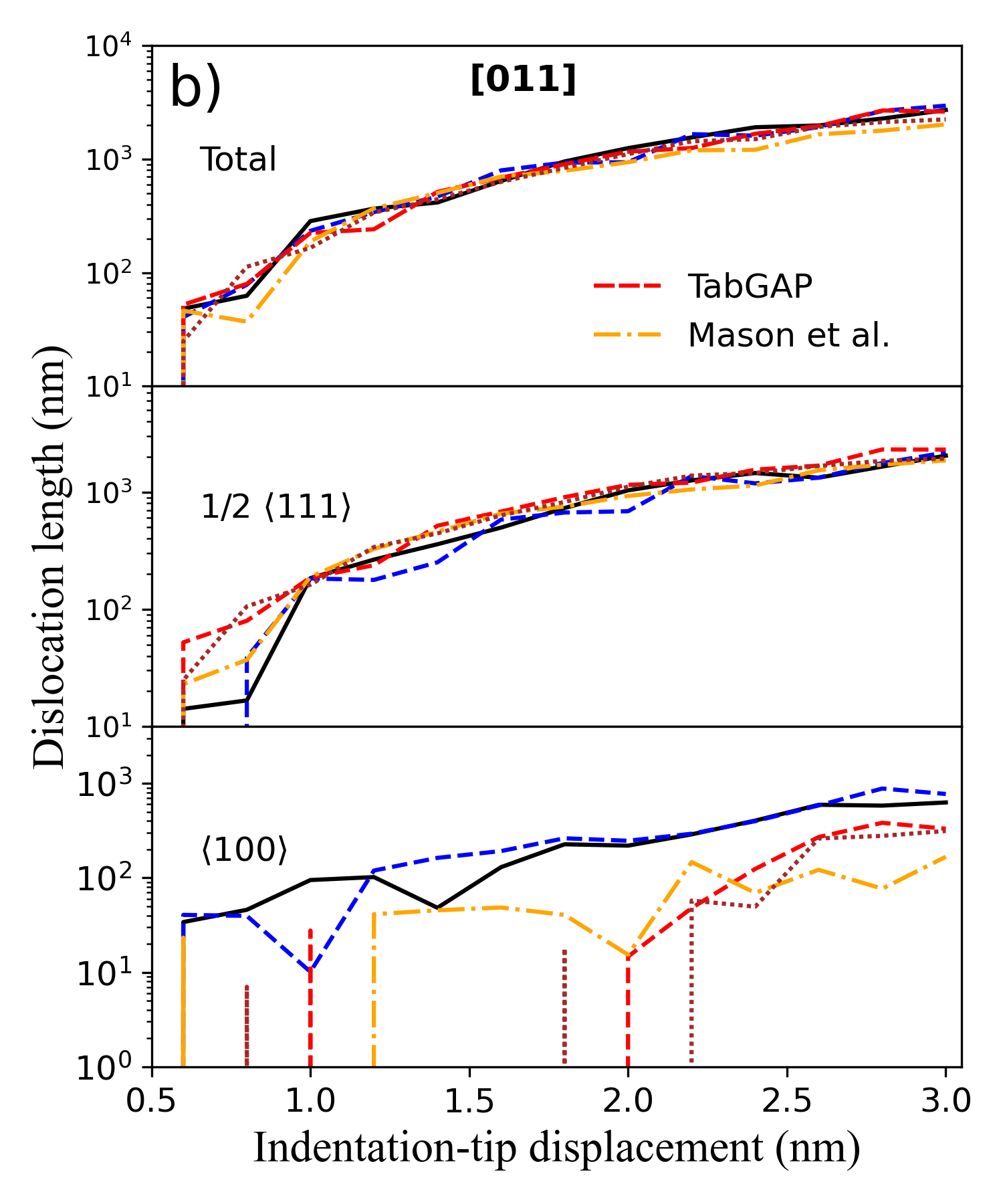}
   \includegraphics[width=0.3\textwidth]{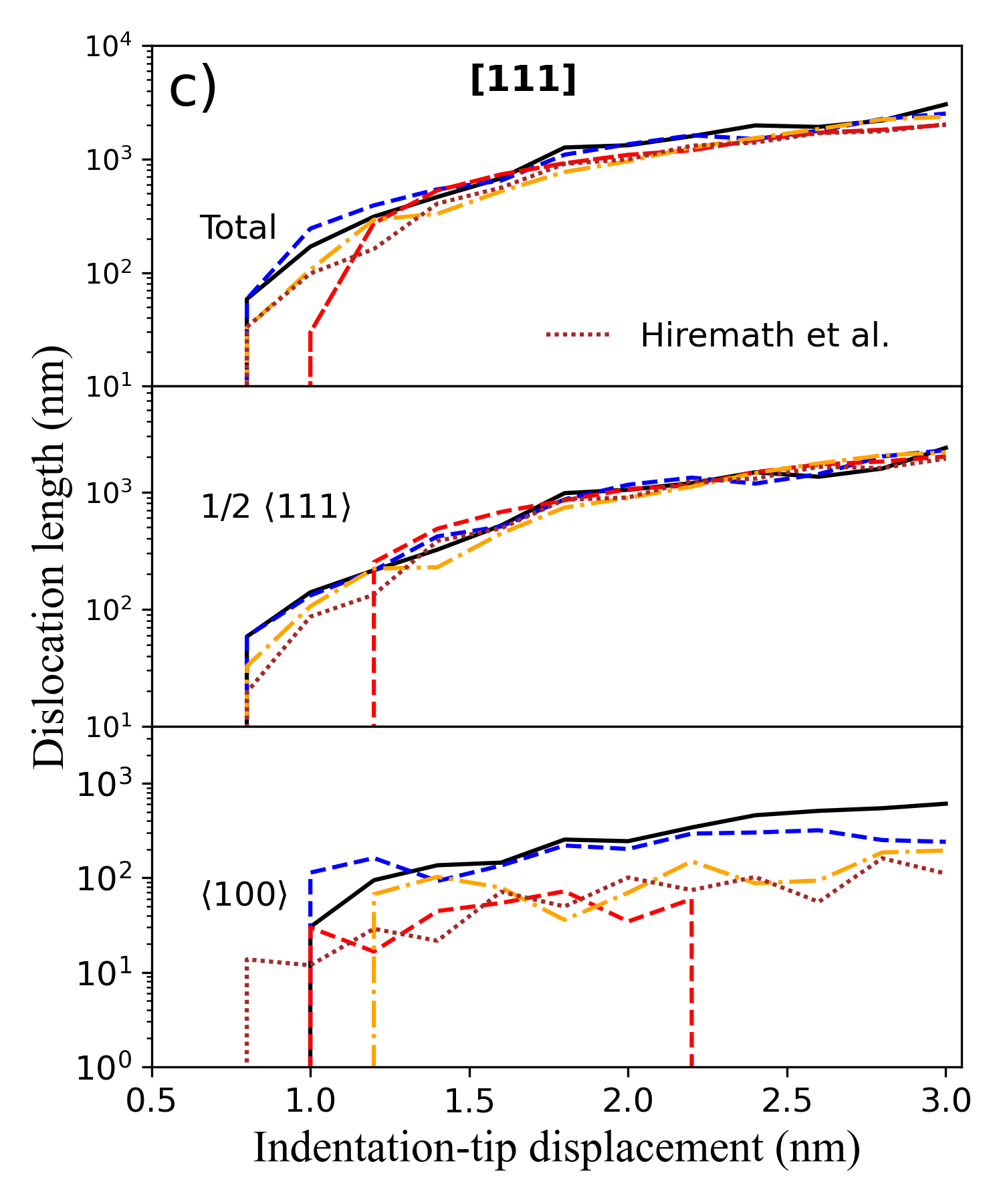}
   \caption{(Color on-line). Total, $1/2 \langle 111 \rangle$ type, 
   and $\langle 100 \rangle$ type dislocation density as a function of
   the indentation depth during loading process for [001] in a), 
   [011] in b), and [111] in c) crystal orientations; showing the 
   differences of the description of W-W interaction by the MD potentials.
   }
   \label{fig:distdens}
\end{figure*}

where $R$ is the indenter radius, $h$ is the indenter 
displacement, and $E_{\rm eff}$ is 
the effective elastic modulus.
Thus, the elastic to plastic deformation transition 
can be identified during the loading process by the 
pop--in event that is well modeled by all the approaches; 
which can be correlated to experimental results to 
study stress distributions under the tip which is in a 
good qualitative good agreement to the data reported by Beake et al.
\cite{BEAKE201863,1998562}.
In our MD simulations, the elastic unloading relaxation is performed 
for long time allowing W atoms to be chemically active 
and mobile under all interaction forces during the elastic and 
plastic relaxation of the material.
%The reduced Young's modulus computed by Oliver Pharr method is 
%173.78 GPa and 172.65 GPa, for the EAM1 and EAM2 potentials,
%respectively.

The effect of the crystal orientation is observed by identifying 
the pop-in event where a sequence is found by every approach used 
following the characteristic maximum pop--in 
load for [001] orientation and the minimum one for [111] 
orientation for BCC metals \cite{Javier2021,VARILLAS2017431,GOEL2018196}.
For the indentation mark left by the tip, tabGAP and MEAM 
potentials report similar trends.
% where is due to the 
%the contribution many body interactions in the 
%computational model and parametrization of the potentials.
All approaches report 
the similar value for the critical pop-in load and effective Young's modulus 
regardless of the crystal 
orientation; albeit of the 
recombination effects observed at $2$ nm depth by the 
MD simulations.

%the Hardness obtained by Oliver-Pharr fitting is similar 
%by all interatomic potentials. 
%Nevertheless, the loading and unloading curves are described in a different 
%way, showing crystallographic effects as well as effects of the interatomic potential. 
The proper description of the nanoindentation test by recorded LD curves 
is of importance in materials design, where W is used for calibration 
of the nanoindenter machine in most of the cases \cite{BEAKE201863}.
Thus, the hardness of the indented sample is 
calculated by applying the Oliver and Pharr method \cite{oliver_pharr_1992} 
following the fitting curve to the unloading process curve as: 
\begin{equation}
P = P_0 \left( h - h_f \right)^m
\label{Eq.OandP}
\end{equation}
with $P$ is the indentation load; 
$h$ is the indentation depth and $h_f$ is the residual depth 
after the whole indentation process; and $P_0$ and 
$m$ are fitting parameters that is added to the unloading curve 
in Fig. \ref{fig:LDcurve}. 
Therefore, the nanoindentation hardness can be computed as: 
$H = P_{\rm max}/A_c$ where $P_{\rm max}$ is the maximum indentation 
load at the maximum indentation depth, $A_c = 
\pi \left(2R - h \right)h_c$ is the 
projected contact area with $R$ as the indenter tip radius which similar 
values by all interatomic potentials.

%Besides, the elastic modulus obtained from the unloading curve is noted 
%to be lower than the effective elastic modulus at first pop-in by Hertz 
%fitting, as reported by our MD simulations and by Goel et al. 
%\cite{GOEL2018196}.

%% file: sections/dislocations.tex
\subsection{\label{sec:results_disl}Dislocation nucleation and evolution}

In order to analyze the dislocation nucleation and evolution of the sample 
during mechanical testing, we identify the different types of dislocations nucleated 
at different indentation depth by using OVITO \cite{ovito} with the DXA package 
\cite{Stukowski_2012}, which provides information of the 
Burgers vector associated to each dislocation. 
Thus, we categorized the 
dislocations into several dislocation types according to their Burgers vectors
as: $1/2 \langle 111 \rangle$, $\langle 100 \rangle$, and 
$\langle 110 \rangle$ dislocation types. 
We compute the dislocation density, $\rho$, as a function of the depth as

\begin{equation}
    \rho = \frac{N_D l}{V_D},
\end{equation}
where $N_D$ is the number of dislocation types, 
$l$ is the dislocation length of each type, and $V_D = 2\pi/3(R_{pl}^{3}-h^{3})$ 
is the volume of the plastic deformation region by using the approximation of a 
spherical plastic zone; where $R_{pl}$ is the largest distance of a dislocation 
measured from the indentation displacement, considering a hemispherical geometry 
, as shown in Fig \ref{fig:distdens}.

In general, dislocation glide occurs in the closest-packed $\langle 
111 \rangle$ directions for BCC metals with Burgers vector 
$b = 1/2 \langle111\rangle$, and slip planes belong to the \{110\}, \{112\}.
To analyze the atomic structure during nanoindentation test 
which provides information about the mechanisms of dislocation 
nucleation and evolution \cite{DeBacker2017}, we compute the 
dislocation density at different indentation depths.
%by using the Dislocation Extraction Analysis (DXA) tool 
%\cite{ovito,Stukowski_2012}. 
The output data is reported in Fig. \ref{fig:distdens} that 
provides the total, $1/2 \langle 111 \rangle$ type, and $\langle 100 \rangle$
type dislocation density as a function of the indentation depth during 
loading process for [001] in a),  [011] in b), and [111] in c) crystal 
orientations. 
All MD simulations results show oscillations of the 
dislocation length during the loading process 
for the $\langle 100 \rangle$ dislocation junction, and the identified 
drop minimum at 2 nm depth for the total length is related 
to the nucleation of a single prismatic loop. 
Note that all potentials are different for 
the nucleation and evolution of the dislocation junction, 
where information and good representation of the surface 
energy and its interaction with the indenter tip 
leads to the mechanism of this junction nucleation. 
We note that this kind of defect is common underneath 
the indenter tip where the interaction or dissociation 
of $1/2 \langle 111 \rangle$ dislocation leads to 
dislocation junction nucleation. 
%The tabGAP approach seems to model this mechanism properly. 
%due to its parametrization to DFT results.

\begin{figure}[!t]
   \centering
  \includegraphics[width=0.375\textwidth]{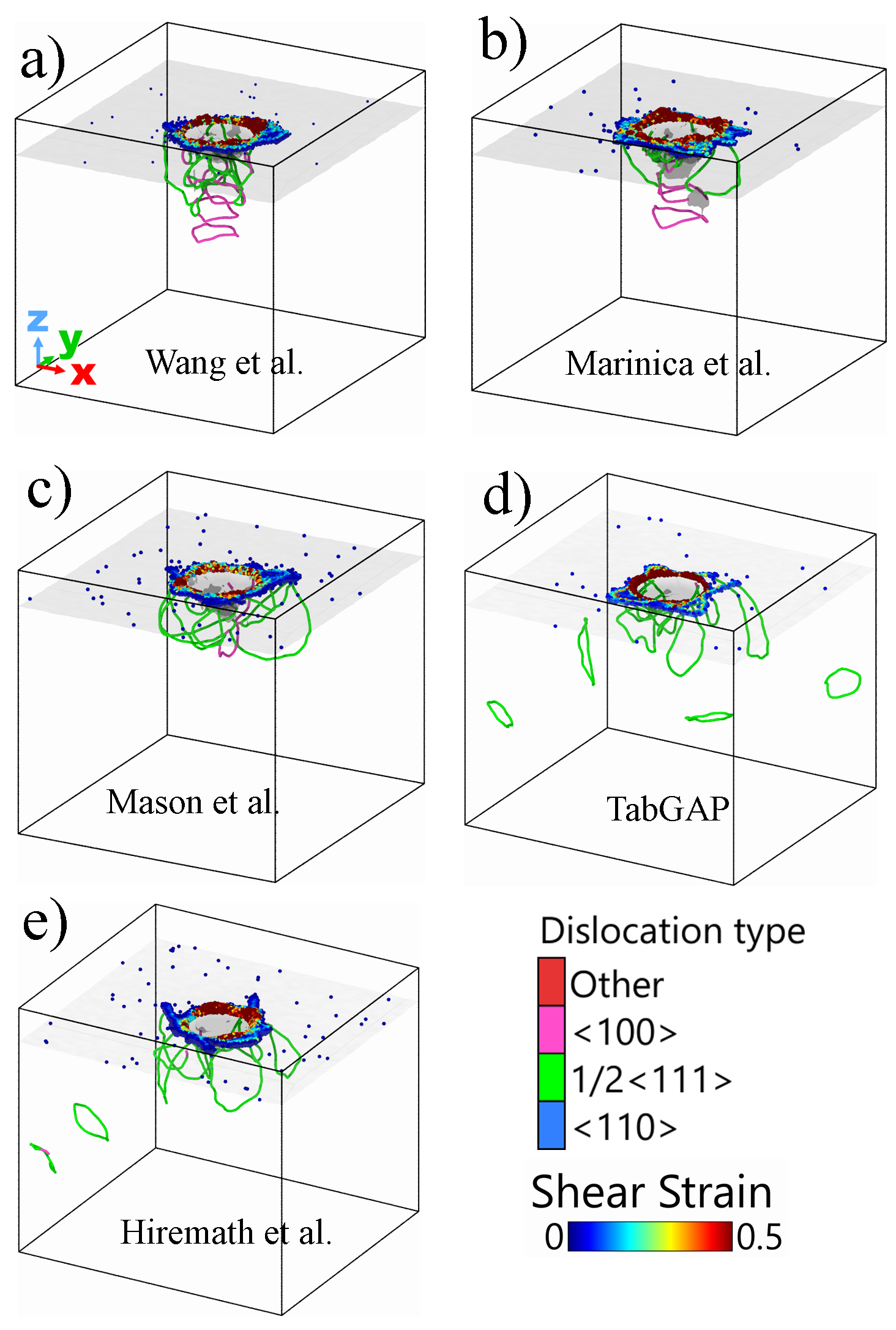}
   \caption{(Color on-line). Visualization of the dislocation 
   nucleation and evolution at different indentation depths by all approaches. 
   A shear dislocation loops is 
   nucleated at the maximum indentation depth by all potentials.
   }
   \label{fig:visdisl}
\end{figure}

It is noted that all the potentials model the dynamics of
the elastic to plastic deformation transition differently,
as shown in Fig \ref{fig:visdisl}. 
Here, we visualize the dislocation network 
nucleated at the maximum indentation depth  for the [001] 
crystal orientation, 
as well as the atomic shear strain mapping that shows pile-ups formation 
and slip traces due to indentation with a maximum strain for region around 
the indenter tip.
This is the main difference between the MD simulations 
performed by every approach. 

The MD simulations performed by using Wang et al. (Fig. \ref{fig:visdisl}a)
and Marinica et al. (Fig. \ref{fig:visdisl}b) potentials 
show similar results where $\langle 100 \rangle$ prismatic loops are 
nucleated.
This is likely due to $\langle 100 \rangle$ loops being more stable than 
1/2$\langle 111 \rangle$ loops in these potentials~\cite{byggmastar_collision_2019}.
%due to the lack of information for the surface energy and 
%interface modeling for material and vacuum. 
In contrast, in the MD simulations with the potential by Mason et al.,
%show the contribution of ZBL corrections 
%to the EAM approach
1/2$\langle 111 \rangle$ half loops are nucleated 
following the slip families (Fig. \ref{fig:visdisl}c). 
Moreover, tabGAP (Fig. \ref{fig:visdisl}d) and Hiremath et al 
(Fig. \ref{fig:visdisl}e) report similar results for the dislocation 
nucleation and dynamics where 1/2$\langle 111 \rangle$ prismatic 
loops follow the [110] slip system and symmetrical one; which is 
expect to observe in the experiments \cite{BEAKE201863,1998562}.
%The [011] crystal orientation, at a depth of $0.5$ nm, 
%a half loop is nucleated by all the potentials
%with a burger vector $1/2 \langle 111 \rangle$ where EAM 
%potentials mainly nucleate a $\langle 100 \rangle$ junction. 
%In addition, for the [111] orientation, 
Besides, the lasso--like dislocation 
is nucleated by all the approaches where a single prismatic loop 
is nucleated at a range of 1.0 to 
1.5 nm, depending on the approach. 
The visualization of the dislocation dynamics during 
nanoindentation test for [011] and [111] orientations are reported
in the supplementary material.

\begin{figure*}[!t]
   \centering
  \includegraphics[width=0.95\textwidth]{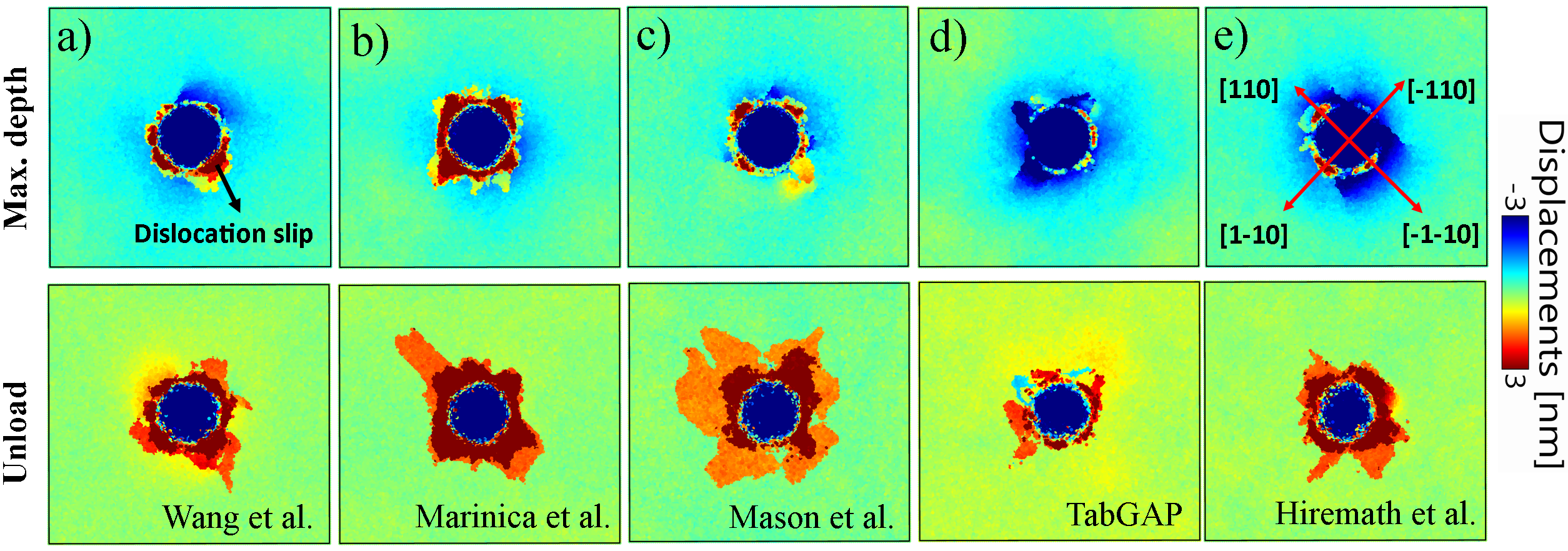}
   \caption{(Color on-line). Visualization of the formation of pile-ups 
   and slip traces for the indented [001] W sample at the maximum depth and after removing 
   the indenter tip.
   }
   \label{fig:pileups}
\end{figure*}

The formation of slip traces at the surface is of importance to
manifest crystal
plasticity process during nanoindentation test where the traces are
indicative of the underlying dislocation glide processes occurring 
in the subsurface, as depicted in Fig. \ref{fig:pileups} for the 
maximum depth 
(upper panel) and after unloading (lower panel) of [001] W sample. 
Due to the interface between the W sample and vacuum, the interatomic 
potentials are required to describe properly the interaction of 
terraces during 
the development of plastic hillocks that exhibits a marked directionality
as a function of crystalline orientation, as do slip traces. 
Although the rosettes presented by MD simulations show the typical 
four-fold one for BCC metal on the [100] orientation as reported 
experimentally 
for W \cite{1998562} and MD simulations for BCC Fe \cite{VARILLAS2017431}, 
the displaced atoms are in different arrangements;
%Twinning is also 
%represented in different manners 
showing the need of accurate modeling for open boundaries simulations.
%MD simulations
%for modeling mechanical testings where open boundaries are considered in
%the simulations.

During the loading process, there is a competition between screw and 
edge glide 
processes of the dislocation loops at the surface that is modeled 
in different way by the chosen MD potential, leading to a couple of 
distinguishable patterns in terms of slip trace formation, as 
presented at the upper panel of Fig. \ref{fig:pileups}. 
The observed correspondence between the nucleated loops and the propagation 
of slip traces can only be understood by considering the associated $\langle 
111 \rangle$ dislocation glide direction is directed towards the surface plane.
From our MD simulations, the formation of indentation plastic imprints is a
process associated to the onset of plastic bursts during mechanical loading 
that is modeled in similar manner by tabGAP and the modified EAM potential, 
where our results are in good agreement 
with experimental SEM and AFM images \cite{1998562,Wangexp}.

\begin{figure}[b!]
    \centering
    \includegraphics[width=0.45\textwidth]{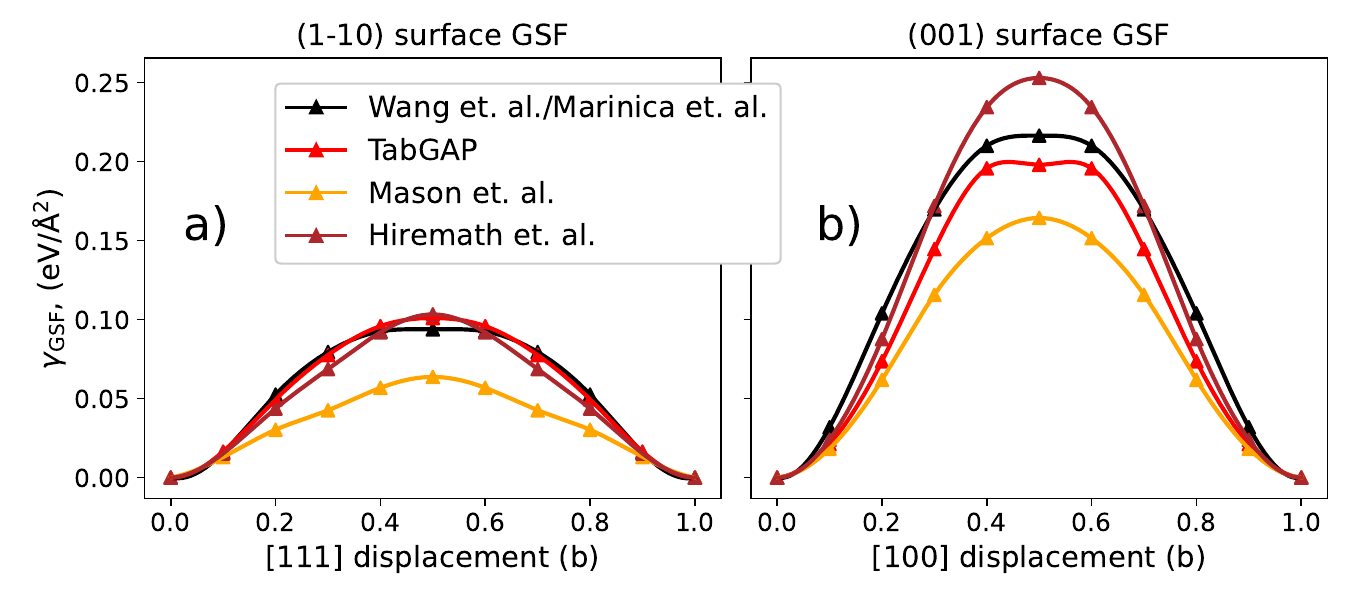}
    \caption{Generalized stacking fault energies for $\langle111\rangle\{110\}$ in a)
    and $\langle100\rangle\{011\}$ in b) computed for all approaches.}
    \label{fig:SFE}
\end{figure}

Characteristically, the stacking fault energy scales as $G\cdot b$, 
while dislocation energies as $G\cdot b^2$, suggesting that a 
fundamental difference in $b$ should influence and amplify effects 
that distinguish dislocations from stacking faults during mechanical
deformation simulations. 
Fig. \ref{fig:SFE} shows results for the generalized staking fault 
energies that are computed by 
defining the $\gamma$--line by cutting a perfect crystal with two 
free surfaces and displacing the two parts relative to each other 
on the chosen direction parallel to the cut plane; with 
periodic boundary conditions applied along the cut plane. 
For each displacement vector, the atomic positions are relaxed only 
in the direction perpendicular to the cut plane
\cite{doi:10.1080/14786431003668793}.
%As we vary the fault vector, the excess energy
%generates a surface, which represents the energies of 
%‘generalized’ stacking faults.
Using unit cells from dislocation objects with $2\times2$ l.u. 
surface area and 30 l.u. perpendicular to the cut plane and 
a force tolerance for relaxation of 0.01 eV/\AA{}.
These results confirmed the different shapes of the slip traces 
showing that Mason et al potetentials sub-estimate the 
SFEs.

%% file: sections/sizeEffects.tex
%%%%%%%%%%%%%%%%%%%%%%%%%%%%%%%%%%%%%
\subsection{Nanoindentation size effects}
\label{subsec:sizeeffec}

\begin{figure}[t!]
    \centering
    \includegraphics[width=0.48\textwidth]{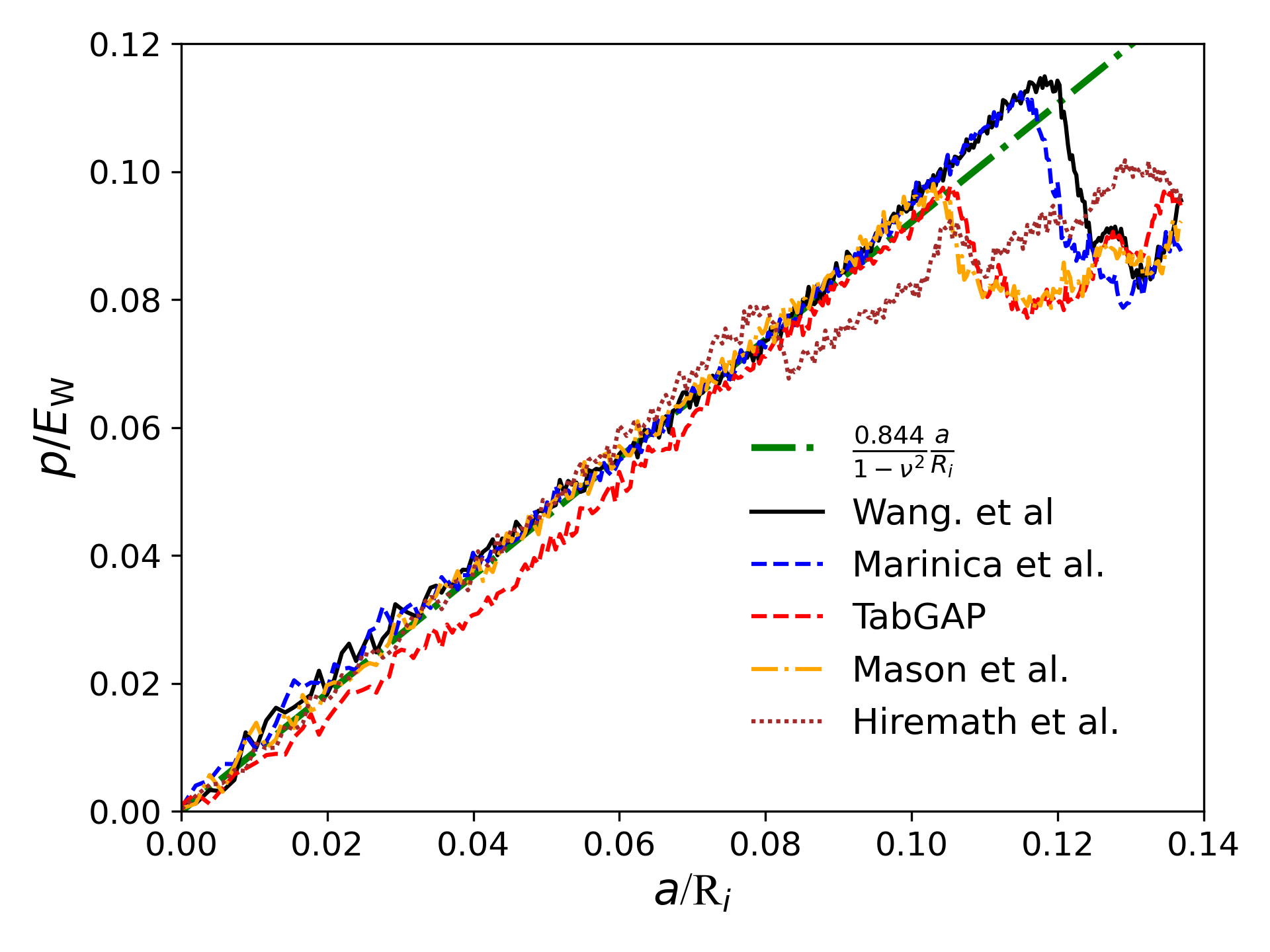}
    \caption{Evolution of normalized contact pressure, $p/E_{\rm W}$, 
    with normalized contact radius, $a/R_i$ ( $R_i = 25$ nm). 
    It is noted that the early elastic process follows the scaling law:
    $0.844/(1-\nu^2)$$a/R_i$.}
    \label{fig:contact_pressure}
\end{figure}

In order to obtain more information about the nanomechanical response during
loading, we perform MD simulation by considering an indenter tip radius $R = 
25$ nm to calculate the indentation stress and strain \cite{JavVarilla};
by taking into account the contact radius between the
sample and the tip by using the geometrical relationship $a(h) =
\sqrt{R_i^2-(R_i-h)^2 }$. 
This expression allows us to write the contact pressure as a function 
of the depth as: $p(h) = P(h)/\pi a(h)^2$, where $P$ is the indentation 
load.
Fig. \ref{fig:contact_pressure} shows the evolution of the contact
pressure normalized to the Young’s modulus ratio of each approach
as: $p/E_{\rm W}$, as a function of the normalized contact radius, 
$a/R_i$. 
The results seem to follow the universal linear relationship as
\cite{JavVarilla}: $0.844/(1-\nu^2)a/R_i$. 
Here the elastic to plastic deformation transition is observed to 
happen at different "nanoindentation strain". 

\begin{figure}[t!]
    \centering
    \includegraphics[width=0.42\textwidth]{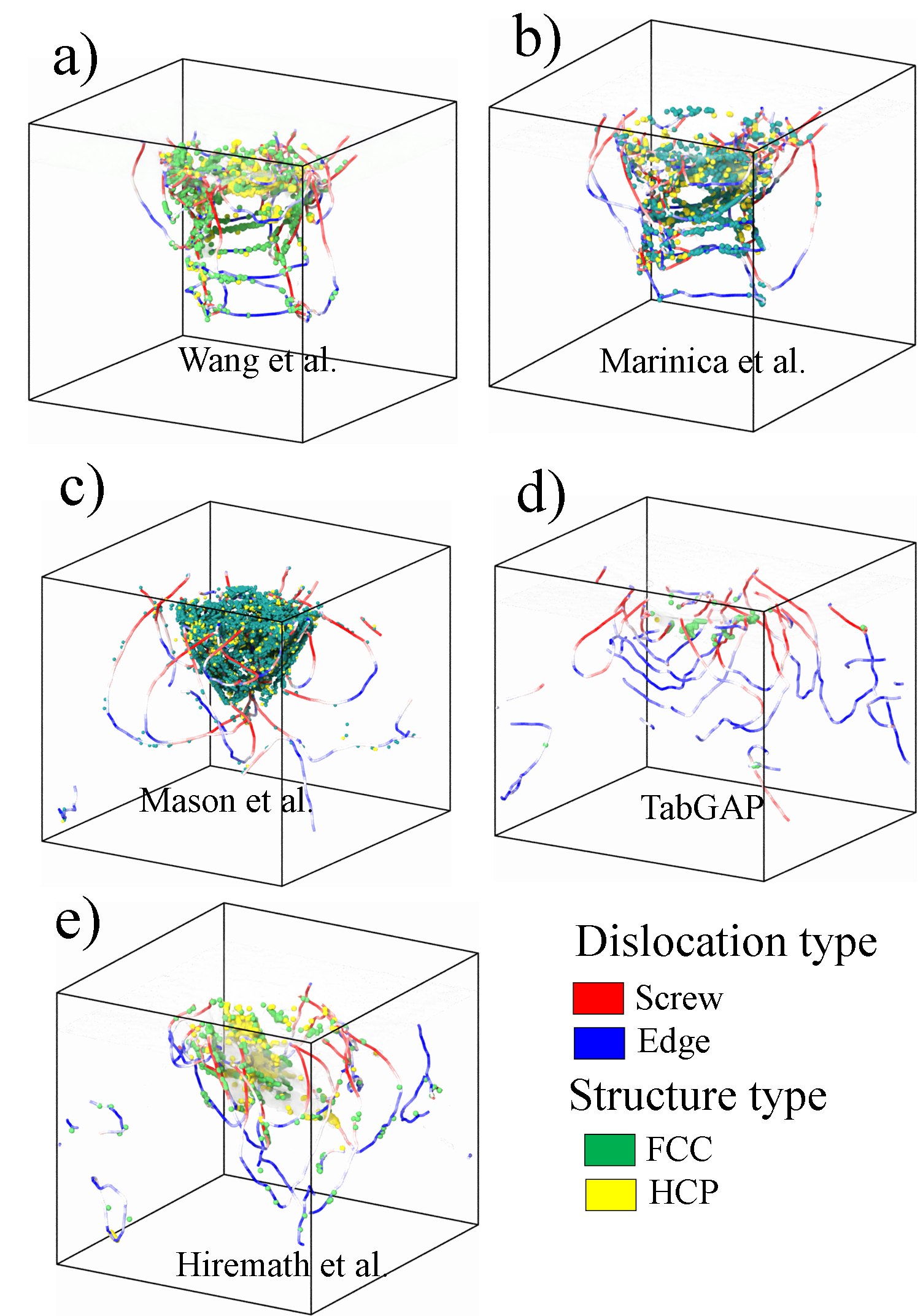}
    \caption{(Color on-line). Dislocation network at maximum depth for an 
    indenter tip radius of 25nm. Dislocation type: screw and edge, as 
    well as structure atom type are shown for all potentials}
    \label{fig:screw_edge}
\end{figure}

In Fig. \ref{fig:screw_edge}, we present the dislocation network at the 
maximum indentation depth by identifying the screw (red lines) and 
edge (blue lines) dislocation for all the approaches. 
In addition, we include the atoms identified as FCC and HCP underneath 
the indenter tip as modeled by different methods.
Results show the dislocation evolution in a larger scale than 
thos presented in Fig. \ref{fig:visdisl}, noticing that screw dislocations 
are mainly located into the plastic region which is underneath the tip. 
While, edge dislocation are mainly distributed deep inside the sample.
An important discrepancy is observed during the nucleation of 
atomic defects; TabGAP and modified EAM potentials tend to nucleate 
the minimum number of atoms, and EAM+ZBL overstimates phase transition 
and nucleation of FCC atoms due to contact pressure.
This is a key difference due to variations on the 
Burgers vector that is defined for BCC crystals as $\vec b = a_0\sqrt{3}/2$, 
it is clear that the potentials are expected to have distinct effects 
on plastic deformation features, such as stacking fault formation 
and dislocation proliferation.

\input{sections/dislocationGlide}

A final remark on the use of different interatomic potentials 
is the computational time and resources that need to be utilized 
for performing MD simulations of nanoindentation test. 
Thus, the lammps software was used in the institutional linux cluster 
with 120 Intel(R) Xeon(R) CPU E5-2680 v2 processors at 2.80GHz with 
computer wall time of 269 mins for Marinica et al;  
165 mins for Wang et al.; 63 mins for Mason et al. (EAM+FS); 
1569 min for Hiremath et al. (Modified EAM); and 
3315 mins for tabGAP. 
Although the tabGAP simulations are faster than the original GAP 
approach, it is considerable slower than EAM based simulations 
and provides a better modeling for mechanical testing, 
as shown in this paper.

%% file: sections/dislocationGlide.tex
%\subsection{Dislocation glide in crystalline W}
%\label{sec:discussion}
%
It is well known that dislocation glide in BCC metals is 
mainly governed by the Peierls barrier, which measures the 
stress that needs to be applied in order to move a dislocation
core to the next atomic valley in the glide plane. 
Thus, the ‘lasso‘ mechanism, as mentioned, is observed by all
methods, suggesting that the main dislocation nucleation
mechanism remains analogous to other BCC metals \cite{JavVarilla,DOMINGUEZGUTIERREZ2021141912}.
Moreover, the Peierls barrier is smaller for edge dislocations
than for screw dislocations, where the BCC metal plasticity
is dominated by the sluggish glide of screw dislocation segments,
as shown by our MD simulations \cite{PhysRevMaterials.4.023601,PhysRevMaterials.2.073608}.

\begin{figure}[t!]
    \centering
    \includegraphics[width=0.45\textwidth]{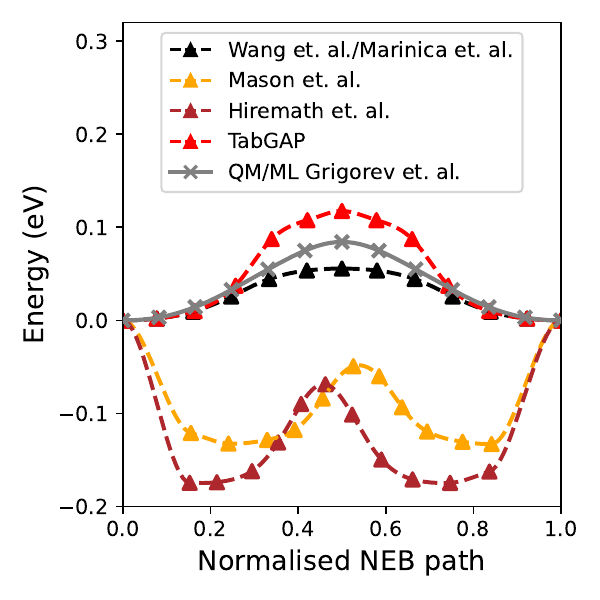}
    \caption{(Color on-line). Screw dislocation glide 1/2(111)\{110\} energy 
    of crystalline W by NEB method for different MD potentials. 
    We compare to reported results by QM/ML calculations \cite{PhysRevMaterials.4.023601}.}
     \label{fig:screwglide}
\end{figure}

\begin{figure}[!b]
    \centering
    \includegraphics[width=0.48\textwidth]{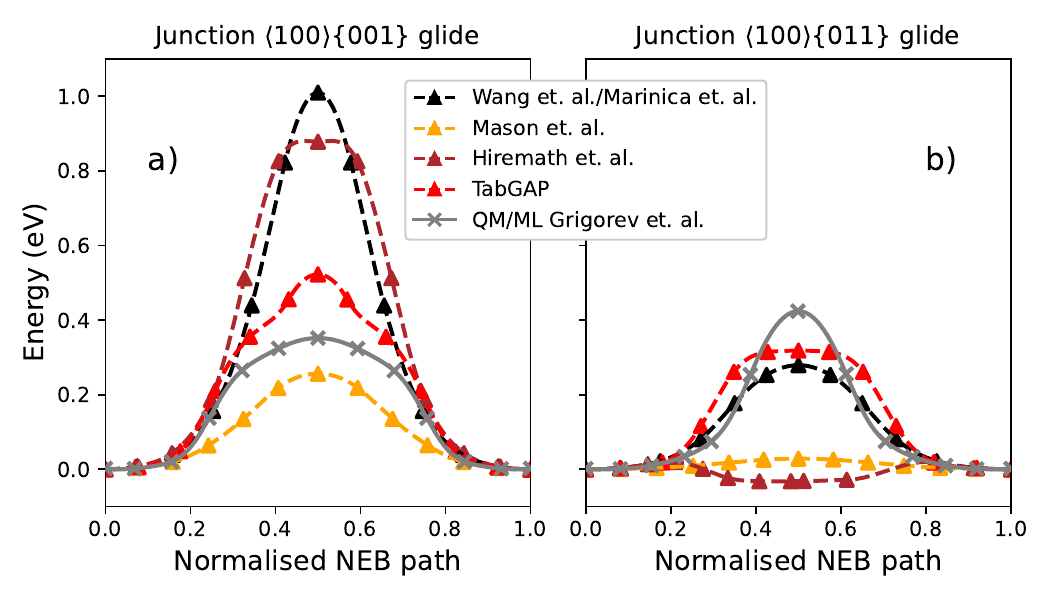}
    \caption{(Color on-line).Junction (100)\{001\} in a) 
              and (100)\{110\} in b) edge
                 dislocation glide barrier in pure tungsten.}
     \label{fig:junctionglide}
\end{figure}

We calculate the screw dislocation glide barriers between two 
easy cores by means of the NEB method with a force tolerance of 
0.025 eV/$\rm \mathring A$ used in  the minimization.
Initial path with 11 intermediate images were obtained by linear
interpolation of the atomic positions between initial and final 
configurations that were relaxed with force 
tolerance of 0.001 $\rm{eV/\mathring A}$. 
The starting dislocation configurations for relaxation were
obtained with anisotropic elasticity method within the Stroh 
formalism \cite{doi:10.1080/14786435808565804} using the elastic 
constants reported in table~\ref{tab:EC_Mo}.
%QM/ML data is taken from [[@Grigorev2021]] - this could probably go to methods. I could add the references to the .bib file, but happy to do it when I have access.
As presented in Fig. \ref{fig:screwglide} and \ref{fig:junctionglide}, 
tabGAP seems to have overall best compromise: comparable barriers 
for two junction dislocations glide planes as well as reasonable 
screw dislocation barriers. 
All other potentials get at least something mislead: a) Hiremath et al. 
and Mason et al. potentials poorly represents screw dislocation core stability. 
%However I can recalculate the barrier taking this into account i.e. the barrier between two split cores. I think it then will be more reasonable.
b) Marinica et al. and Hiremath et al. significantly overestimate the glide 
barrier for a junction dislocation in the \{001\} glide plane; and b) Hiremath 
et al. and Mason et al. predict almost zero glide for the junction dislocation
in the \{011\} glide plane. 
The obtained results can guide experiments
on the understanding of the fundamental mechanisms
for dislocations nucleation with in situ transmission 
electron microscopy
(TEM) images \cite{doi:10.1021/acsami.7b11103}.

%\begin{figure}[!b]
%    \centering
%    \includegraphics[width=0.48\textwidth]{Figures/stackingFault.png}
%    \caption{(Color on-line).Junction (100)\{001\} in a) 
%              and (100)\{110\} in b) edge
%                 dislocation glide barrier in pure tungsten.}
%     \label{fig:junctionglide}
%\end{figure}

%% file: sections/concluding.tex
\section{\label{sec:concluding}Concluding remarks}

In this work, MD simulations by using  different %EAM-based 
interatomic potentials like traditional EAM potentials, a 
recently developed tabulated Gaussian 
approximation potential and a 
modified EAM approaches are performed to investigate 
the nanomechanical response of crystalline W during
nanoindentation tests for [001], [011], 
and [111] orientations. 
We analyzed the dislocation nucleation
and evolution mechanisms described by all approaches.  
We characterized the nanoindentation process in W by tracking 
shear strain accumulation and displacement atoms mapping that can 
be compared to SEM investigations.
Our work is summarized as follows: 1) the comparison between 
the potentials report similarities for the recorded load 
displacement curves, as well as the rosettes formed at the 
surface projections 
of the preferential gliding direction during nanoindentation; 
2) dislocation nucleation mechanisms are 
modeled in a different way during nanoindentation test
due to difference of the Burgers vector magnitude related
to stacking fault formation and dislocation
dynamics. Although a prismatic loop is nucleated 
by both approaches, the dynamics is modeled in 
different manners, where tabGAP 
and MEAM report similar results.
Based on the present results, we conclude that nanomechanical 
tests can be modeled by several interatomic potentials where 
the load displacement and stress--strain curves can be similar. 
However, dislocation dynamics depends on the approach used
to developed the MD potentials exhibit by NEB calculations of 
screw and edge dislocation glide of 1/2$\langle 111 \rangle \{110\}$ and Junction
$\langle 100 \rangle \{001\}$ 
and $\langle 100 \rangle \{110\}$ energies, and stacking fault 
energies showing that tabGAP simulations can emulate 
nanoindentation test as close as possible to experiments. 
In our future work, we will investigate the nanomechanical 
response of chemically complex BCC metals under external load 
by using recently developed tabGAP potentials that can 
be compared to experimental SEM and TEM images.
%These results also open an opportunity to study dislocation 
%nucleation mechanism in hydrogenated W samples.